\documentclass[11pt]{article}
\usepackage{amsmath,amssymb,color,graphics,epsfig}
\usepackage{amsfonts}
\newcommand{\hoch}[1]{$\, ^{#1}$}

\newcommand{\be}{\begin{equation}}
\newcommand{\ee}{\end{equation}}
\newcommand{\bea}{\setlength\arraycolsep{2pt} \begin{eqnarray}}
\newcommand{\eea}{\end{eqnarray}}

\def\ft#1#2{{\textstyle{\frac{\scriptstyle #1}{\scriptstyle #2} } }}
\def\fft#1#2{{\frac{#1}{#2}}}

\def\0{{\sst{(0)}}}
\def\1{{\sst{(1)}}}
\def\2{{\sst{(2)}}}
\def\3{{\sst{(3)}}}
\def\4{{\sst{(4)}}}
\def\5{{\sst{(5)}}}
\def\6{{\sst{(6)}}}
\def\7{{\sst{(7)}}}
\def\8{{\sst{(8)}}}
\def\sst#1{{\scriptscriptstyle #1}}

\begin{document}

\begin{flushright}
%\hfill{KIAS-P12028}
 %\hfill{
%\bf hep-th/yymmnnn}
\end{flushright}

\vspace{25pt}
\begin{center}
{\large {\bf Scalar Charges in Asymptotic AdS Geometries}}

\vspace{10pt}
Hai-Shan Liu\hoch{1} and H. L\"u\hoch{2}

\vspace{10pt}

\hoch{1} {\it Institute for Advanced Physics \& Mathematics,\\
Zhejiang University of Technology, Hangzhou 310023, China}

\vspace{10pt}

\hoch{2}{\it Department of Physics, Beijing Normal University,
Beijing 100875, China}

\vspace{40pt}

\underline{ABSTRACT}
\end{center}

We show that for $n$-dimensional Einstein gravity coupled to a scalar field with mass-squared $m_0^2=-n(n-2)/(4\ell^2)$, the first law of thermodynamics of (charged) AdS black holes will be modified by the boundary conditions of the scalar field at asymptotic infinity.  Such scalars can arise in gauged supergravities in four and six dimensions, but not in five or seven. The result provides a guiding principle for constructing designer black holes and solitons in general dimensions, where the properties of the dual field theories depend on the boundary conditions.

\vfill {\footnotesize Emails: hsliu.zju@gmail.com\ \ \ mrhonglu@gmail.com}

\thispagestyle{empty}

\pagebreak
%\voffset=0pt
%\setcounter{page}{1}

%\tableofcontents
%\addtocontents{toc}{\protect\setcounter{tocdepth}{2}}

%%%%%%%%%%%%%%%%%%%%%%%%%%%%%%%%%%%%%%%%

\newpage
%%%%%%%%%%%%%%%%%%%%%%%%%%%%%%%%%%%%%%%%

{\bf Introduction}: Black holes are one of the most important classes of object predicted by Einstein's theory of gravity.  Many important properties of black holes have been studied and established.  Black holes that are asymptotic to anti-de Sitter spacetime (AdS) are particularly useful in the AdS/CFT correspondence \cite{ads-cft1}, for studying quantum field theory or even condensed matter physics at finite temperature. Numerical evidence suggests that black holes can develop scalar ``hair'' in both asymptotic AdS or flat spacetimes \cite{Sudarsky:2002mk}. Recently, many explicit examples of scalar hairy black holes in four and higher dimensions have been found \cite{commongrave}.  This leads to a natural question as to whether there exist scalar charges, analogous to those associated with electromagnetic fields.  The subject of scalar charge and its contribution to the first law of black hole thermodynamics is not well understood.

If a theory with a scalar $\phi$ has a global symmetry with a constant shift, $\phi\rightarrow \phi+ c$, as in the case of the dilaton in ungauged supergravities, the asymptotic value $\phi_0$ at infinity is an integration constant, and can be treated as a scalar charge \cite{Gibbons:1996af}.  If the scalar has a potential with a stationary point $\phi=\phi_0$, the global symmetry disappears and $\phi$ takes the {\it fixed} value $\phi_0$ at infinity.  In this situation, there is no obvious definition of a scalar charge and one might expect that the first law would not be modified by the contribution from the scalar. In four dimensions, it was shown that a massless scalar (in the AdS sense) with the large-$r$ boundary behavior
\be
\phi=\fft{\phi_1}{r} + \fft{\phi_2}{r^2} + \cdots
\ee
can break some of the boundary AdS symmetries unless one of the following three conditions is satisfied: (1) $\phi_1=0$, or (2) $\phi_2=0$ \cite{bf}, or (3) $\phi_2/\phi_1^2$ is some fixed constant \cite{Hertog:2004dr}. (See also \cite{Henneaux:2002wm}.) Solitons that violate all these three conditions were constructed numerically in \cite{Hertog:2004ns}, giving rise to ``designer gravity'', where the properties of the field theory depend on the boundary conditions. In \cite{Lu:2013ura}, a Kaluza-Klein dyonic AdS black hole in four-dimensional maximal gauged supergravity was constructed, for which the scalar boundary behavior also violates these three criteria.  The consequence is that the naively-expected first law of thermodynamics $dM=TdS + \Phi_edQ_e + \Phi_p dQ_p$ for the dyon does not hold.  Instead $dM$ in the first law is shifted by a 1-form $-XdY\equiv Z$, given by \cite{Lu:2013ura}
\be
Z=\ft{1}{12\ell^2} (2\phi_2d\phi_1 - \phi_1 d\phi_2)\,,
\ee
where $\ell$ is the radius of the asymptotic AdS.  The first law becomes
\be
dM=TdS + \Phi_edQ_e + \Phi_p dQ_p + X dY\,.
\ee
It reduces to the standard one if any of the three criteria mentioned above is met.

In this paper, we show that this phenomenon can occur also in higher dimensions, and we determine the conditions for a non-vanishing $Z$.  We begin by considering $n$-dimensional Einstein gravity coupled to a scalar field with a generic potential
\be
{\cal L}=\sqrt{-g}\big(R- \ft12 (\partial\phi)^2 - V(\phi)\big)\equiv\sqrt{-g} L_0\,.\label{theory1}
\ee
The equations of motion are
\be
E_{\mu\nu}\equiv R_{\mu\nu} - \ft12\partial_\mu\phi\partial_\nu\phi -\ft1{n-2} V g_{\mu\nu}=0\,,\qquad \Box\phi = \fft{dV}{d\phi}\,.\label{purescalareom}
\ee
Many explicit examples of exact solutions for hairy black holes have been obtained for some specific choices of the scalar potential \cite{commongrave}.  We would like to examine whether these solutions admit a non-vanishing $Z$.

{\bf Wald's canonical charge}: We begin by reviewing Wald's formalism for deriving the first law of thermodynamics by the Noether procedure \cite{Wald:1993nt}.  We first consider a generic variation of the Lagrangian (\ref{theory1}):
\be
\delta {\cal L} = {\rm e.o.m.} + \sqrt{-g}\, \nabla_\mu J^\mu\,,\label{variation1}
\ee
where e.o.m.~denotes the equations of motion for the fields, and
\be
J^\mu = g^{\mu\rho}g^{\nu\sigma} (
\nabla_{\sigma}\delta g_{\nu\rho} - \nabla_{\rho}\delta g_{\nu\sigma}) - \nabla^\mu \phi \delta\phi\,.
\ee
From this one can define a 1-form $J_\1=J_\mu dx^\mu$ and its Hodge dual $\Theta_{\sst{(n-1)}}=(-1)^{n+1}{*J_{\1}}$.  We now specialize to a variation that is induced by an infinitesimal diffeomorphism $\delta x^\mu=\xi^\mu$.  One can show that
\be
J_{\sst{(n-1)}}\equiv \Theta_{\sst{(n-1)}} - i_{\xi} {*L_0} = {\rm e.o.m.} -
d{*J_\2}\,,
\ee
where $i_\xi$ denotes a contraction of $\xi^\mu$ on the first index of the $n$-form ${*L_0}$, and $J_\2=d\xi_\1$ with $\xi_\1=\xi_\mu dx^\mu$.
One can thus define an $(n-2)$-form $Q_{\sst{(n-2)}}\equiv {*J_\2}$, such that $J_{\sst{(n-1)}}=dQ_{\sst{(n-2)}}$.  Note that we use the subscript notation ``$(p)$'' to denote a $p$-form. To make contact with the first law of black hole thermodynamics, we take $\xi^\mu$ to be the time-like Killing vector that is null on the horizon.  Wald shows that the variation of the Hamiltonian with respect to the integration constants of a given solution is \cite{Wald:1993nt}
\be
\delta H=\fft{1}{16\pi}\delta \int_c J_{\sst{(n-1)}} - \int_c d(i_\xi \Theta_{\sst{(n-1)}}) =\fft{1}{16\pi}\int
_{\Sigma^{(n-2)}} \Big(\delta Q_{\sst{(n-2)}} - i_\xi \Theta_{\sst{(n-1)}}\Big)\,,
\ee
where $c$ denotes a cauchy surface and $\Sigma^{(n-2)}$ is its two boundaries, one at infinity and one on the horizon.

{\bf Application in Einstein-scalar theory}: We now apply Wald's formalism to the Einstein-scalar theory (\ref{theory1}).  In this paper, we shall mainly consider static and spherically-symmetric solutions. For our purpose, it is convenient to write the metric ansatz in Schwarzschild-like coordinates, with
\be
ds_n^2 = - h(r) dt^2 + \fft{dr^2}{\tilde h(r)} + r^2 d\Omega_{n-2}^2\,,
\ee
where $d\Omega_{n-2}^2$ is the metric on the unit $S^{n-2}$.
Assuming that the metric is well behaved at asymptotic infinity with $h\sim \tilde h\sim \ell^{-2} r^2$ , the properly-normalized time-like Killing vector is $\xi=\partial/\partial t$.  We then find
\be
Q_{\sst{(n-2)}} = r^{n-2} h' \sqrt{\ft{\tilde h}{h}}\, \Omega_{\sst{(n-2)}}\,,
\ee
where a prime denotes a derivative with respect to $r$ and $\Omega_{\sst{(n-2)}}$ is the volume form of the unit $S^{n-2}$.
The quantity $i_\xi \Theta_{\sst{(n-1)}}$ has contributions from both the gravity sector and the scalar sector:
\bea
i_\xi \Theta^{\rm grav}_{\sst{(n-1)}} = r^{n-2} \Big(
\delta\big(h'\sqrt{\ft{\tilde h}{h}}\big) + \ft{n-2}{r} \sqrt{\ft{h}{{\tilde h}}} \delta\tilde h \Big)\Omega_{\sst{(n-2)}}\,,\quad
i_\xi \Theta^{\phi}_{\sst{(n-1)}} = r^{n-2} \sqrt{h\tilde h}\, \phi' \delta\phi
\Omega_{\sst{(n-2)}}\,.
\eea
It is clear that the scalar contribution vanishes on the horizon where $h$ and $\tilde h$ vanish, and so $\delta H$ evaluated on the horizon is simply $TdS$.  It then remains to evaluate the contributions from the sphere at infinity.
We consider only the case where the mass contributes the leading-order deviation of $g_{tt}$ from the AdS, namely
\be
h=\ell^{-2}r^2 + 1 - \fft{m}{r^{n-3}} + \fft{m_1}{r^{n-2}} + \cdots\,.
\label{hfalloff1}
\ee
We find
\be
\delta H=\fft{1}{16\pi}\int_{r\rightarrow\infty} (\delta Q - i_\xi \Theta)
=-\fft{\omega_{n-2}}{16\pi}\lim_{r\rightarrow \infty} r^{n-2} \Big(\ft{n-2}{r} \delta \tilde h+\sqrt{h\tilde h}\,\phi'\delta\phi\Big)
= \delta M +Z\,,
\ee
where
\be
M=\fft{(n-2)\omega_{n-2}}{16\pi}\, m\,,\qquad
Z=- \fft{\omega_{n-2}}{16\pi}\lim_{r\rightarrow \infty} r^{n-2} \Big(\ft{n-2}{r} \delta (\tilde h-h) + \ell^{-2}r^2\phi'\delta\phi\Big)\,.\label{mz}
\ee
Here $\omega_{n-2}=\int \Omega_{\sst{(n-2)}}$ and $M$ is the mass. Thus whether the quantity $Z$ diverges, converges or vanishes depend on the specific falloffs of $\phi$ and $\tilde h-h$.  For solutions with no scalar, such as the Schwarzschild or Reissner-Nordstr\o m AdS black holes, $h=\tilde h$ and hence $Z$ vanishes identically. It was shown that the quantity $Z$ for the Kaluza-Klein dyonic AdS black hole in four dimensions is finite and non-vanishing \cite{Lu:2013ura}.  More general solutions involving multiple dyonic charges in the STU gauged supergravity model were constructed in \cite{Chow:2013gba}.  The non-vanishing of $Z$ was interpreted in \cite{Chow:2013gba} as indicating that the mass is not well defined.  We prefer to take the view proposed in \cite{Lu:2013ura}, that one can still give a meaningful definition of mass, but with the first law modified by the addition of $Z=-XdY$.

At the first sight, one may think that the quantity $Z$ could only be evaluated in explicit solutions, where the falloffs of $\phi$ and the metric functions are known.  In fact rather general statements can be made without knowing the explicit solutions.  Two linear combinations of the Einstein equations (\ref{purescalareom}) do not involve the scalar potential. In particular, the combination $E_{t}{}^t - E_{i}{}^i=0$, where $i$ denotes any specific sphere direction, does not involve the scalar at all:
\be
\fft{h''}{h}-\fft{h'^2}{2h^2} +\fft{h'\tilde h'}{2h\tilde h} +
\fft{(n-3)h'}{rh'} - \fft{\tilde h'}{r\tilde h} - \fft{2(n-3)(\tilde h-1)}{r^2 \tilde h}=0\,.\label{htheom}
\ee
Thus we find that for (\ref{hfalloff1}), the leading falloff for $\tilde h-h$ is given by $\fft{\Delta_1}{r^{n-4}}$, where $\Delta_1$ is the integration constant, proportional to $m_1$ in (\ref{hfalloff1}).  From $E_t{}^t-E_{r}{}^r=0$, we have
\be
\phi'^2 = \fft{(n-2)(\tilde h h'- h\tilde h')}{rh\tilde h}\,.\label{phipsquare}
\ee
Thus the leading falloff for $\phi$ is $\fft{\phi_1}{r^{(n-2)/2}}$.
Assuming that the solutions are well defined at infinity and can be expanded as:
\bea
h &=&\ell^{-2} r^2 + 1 - \fft{m}{r^{n-3}} + \fft{m_1}{r^{n-2}} + \cdots\,,\qquad
\tilde h-h = \fft{\Delta_1}{r^{n-4}} + \fft{\Delta_2}{r^{n-3}} + \cdots\,,\label{hthfalloffs}\\
\phi &=& \fft{\phi_1}{r^{(n-2)/2}} + \fft{\phi_2}{r^{n/2}} + \cdots\,,
\label{requiredfalloff}
\eea
we find that $\Delta_1$ and $\Delta_2$ are related to $\phi_1$ and $\phi_2$ by
\be
\Delta_1 = \ft14 \ell^{-2} \phi_1^2\,,\qquad \Delta_2 = \ft{n}{2(n-1)} \ell^{-2} \phi_1\phi_2\,.\label{deltaphi}
\ee
Note that the relations (\ref{deltaphi}) are fixed by equation (\ref{phipsquare}).
The coefficient of the $1/r^{2(n-3)}$ falloff in $h$ turns out always to be zero in the absence of electromagnetic charges. Substituting these asymptotic behaviors into (\ref{mz}), we find
\bea
Z &=&\ft{\omega_{n-2}}{16\pi}\Big( -(n-2)\delta \Delta_2 + \ft12 \ell^{-2} (n\phi_2\delta\phi_1 -(n-2)\phi_1 \delta\phi_2)\Big)\cr
&=& \ft{\omega_{n-2}}{32\pi (n-1)\ell^2} \big(n\phi_2\delta\phi_1 - (n-2)\phi_1 \delta\phi_2\big)\,.
\label{Zres}
\eea
It is important to note that the $\Delta_1$ contribution to $Z$ yields a divergent result, but it cancels precisely the contribution from the leading term from $\phi'\delta\phi$. Thus we conclude that $Z$ is finite, and in general non-vanishing unless either (1) $\phi_1=0$, or (2) $\phi_2=0$, or (3) $\phi_2/\phi_1^{n/(n-2)}$ is a fixed constant.  In four dimensions, we recover the result obtained in \cite{Lu:2013ura}.  Note that if we have multiple scalars in a linear $\sigma$-model, the quantity $Z$ is then simply the summation of the contributions of each scalar as in (\ref{Zres}).  Although we focused on solutions with spherical symmetry, the formula (\ref{Zres}) applies also to AdS black holes with planar horizon geometries.

{\bf Scalar properties}: We define the mass of scalar in the linearized equation in the AdS background as follows
\be
\Big(\Box +\fft{2(n-3)}{\ell^2} - {\cal M}^2\Big)\phi=0\,.
\ee
In this definition, scalars in gauged supergravities are massless.  From the leading falloffs of the scalar in (\ref{requiredfalloff}), we find that
\be
{\cal M}^2 = -\fft{(n-4)(n-6)}{4\ell^2}\,.\label{massdef}
\ee
Thus the required mass of the scalar is zero in four and six dimensions.  A massless scalar has a typical falloff of $1/r^{n-3}$, which coincides with the required falloff (\ref{requiredfalloff}) in $D=4$.  In six dimensions, the other falloff of a massless scalar is $1/r^2$, which coincides also with (\ref{requiredfalloff}).  The required mass (\ref{massdef}) implies
that the leading-order expansions of the scalar potential must be
\be
V=-(n-1)(n-2)\ell^{-2} - \ft18 n(n-2) \ell^{-2} \phi^2 + \cdots\,.
\ee
The ``bare'' mass-squared is thus $m_0^2=- n(n-2)/(4\ell^{2})$, corresponding to a conformally massless scalar. The conformal dimension of the dual (relevant) operator of the asymptotic AdS boundary field theory is therefore $\Delta=n/2$. Interestingly, for this particular conformal dimension, the coefficients $\phi_1$ and $\phi_2$ are the zero modes of the two boundary fields $A(x)$ and $B(x)$ at asymptotic infinity:
\be
\phi(x,r) \sim  \fft{A(x)}{r^{d-\Delta}} + \fft{B(x)}{r^\Delta}\,,\label{phifalloff}
\ee
where $d=n-1$ is the dimension of the boundary field theory. This implies that the falloff coefficients $\phi_1$ and $\phi_2$ in (\ref{requiredfalloff}) coincide with the boundary values of the boundary fields $A$ and $B$.

Since scalars in gauged supergravities are massless, our results imply that they cannot contribute a non-vanishing $Z$ in five and seven dimensions.  In six dimensions, the massless scalar can be embedded in gauged supergravity, but only the solutions with both the slower scalar falloff $1/r^2$ and the typical $1/r^3$ can give rise to non-vanishing $Z$. In four dimensions, the requirement selects the ``normal'' massless scalar, and indeed dyonic AdS black holes of gauged supergravity with a non-vanishing $Z$ were constructed \cite{Lu:2013ura,Chow:2013gba}.  On the other hand, the quantity $Z$ vanishes for all the hairy black holes recently constructed in \cite{commongrave}.

{\bf Charged system:}  We can also add a Maxwell field to the system, with the Lagrangian
\be
e^{-1}{\cal L}_A = -\ft14 e^{a\phi} F^2\,,\qquad F=dA\,.
\ee
The electric ansatz that satisfies both the Bianchi identity and the Maxwell equation is
\be
F=\fft{Q e^{-a\phi}}{r^{n-2}} \sqrt{\ft{h}{\tilde h}}\, dr\wedge dt\,.
\ee
Equation (\ref{htheom}) is now modified by the charge, and it is straightforward to verify that the parameter $Q$ enters first at the $1/r^{2(n-3)}$ level in the large-$r$ expansion of $h$.  However, Equation (\ref{phipsquare}) is unchanged. It is then clear that the relations (\ref{deltaphi}) between $\Delta_i$ and $\phi_i$ ($i=1,2$), which are determined by (\ref{phipsquare}), will not be modified by the charge.  Thus we conclude that the formula (\ref{Zres}) for $Z$ holds also for charged solutions.

{\bf Conclusions:}  In this paper, we obtained the formula for calculating the modification of the first law of thermodynamics for asymptotic AdS black holes
due to scalar charges.  In $n$ dimensions, a scalar with bare mass-squared $m_0^2=-n(n-2)/(4\ell^2)$ falls off at large radius as in (\ref{requiredfalloff}).  We find that the first law will be modified, with $dM$ replaced by $dM + Z$ where $Z$ is given by (\ref{Zres}).  This scalar has the unique property that the two decay modes are separated in order by one inverse power of $r$, and hence can conspire to modify the variation of the Hamiltonian in the Wald formalism.  The dual operator in the boundary field theory is relevant, with a conformal dimension $\Delta=n/2$. In four and six dimensions, such a scalar can arise in gauged supergravities, whilst in five and seven dimensions, it cannot.  Our results provide a guiding principle for constructing new designer black holes and solitons, which can be used to compute certain effective potentials of the dual field theories, whose properties depend on the boundary conditions of the scalar field.

\section*{Acknowledgement}

We are grateful to Chris Pope for useful discussions. H-S.L.~is supported in part by the NSFC grant 11305140 and SFZJED grant Y201329687. H.L.~is supported in part by the NSFC grants 11175269 and 11235003.

\end{document}